\documentclass[prc,aps,floats,showpacs,twoside,superscriptaddress,floatfix,twocolumn]{revtex4}
\usepackage{graphicx,pstricks,epsfig,rotating,amsmath}
\newcommand{\gsim}{\stackrel{\scriptstyle >}{\phantom{}_{\sim}}}

\newcommand{\bea}{\begin{eqnarray}}
\newcommand{\eea}{\end{eqnarray}}
\newcommand{\apgt} {\ {\raise-.5ex\hbox{$\buildrel>\over\sim$}}\ }

\graphicspath{{EsymPaperFigs/}}

\begin{document}
\title{Universal symmetry energy contribution to the neutron star equation of state}
\date{\today}

\author{D.~Blaschke}
\email{blaschke@ift.uni.wroc.pl}
\affiliation{
Institute for Theoretical Physics,
University of Wroc{\l}aw,
Max Born Pl. 9,
50-204 Wroc{\l}aw, Poland}
\affiliation{
Bogoliubov  Laboratory of Theoretical Physics,
Joint Institute for Nuclear Research,\\
Joliot-Curie Street 6,
141980 Dubna,
Russia}
\affiliation{
National Research Nuclear University (MEPhI),
Kashirskoe Shosse 31, 
141980 Moscow,
Russia}

\author{D.E. Alvarez-Castillo}
\email{alvarez@theor.jinr.ru}
\affiliation{
Bogoliubov  Laboratory of Theoretical Physics,
Joint Institute for Nuclear Research,\\
Joliot-Curie Street 6,
141980 Dubna,
Russia}
\affiliation{Instituto de F\'{\i}sica,
Universidad Aut\'onoma de San Luis Potos\'{\i}\\
Av. Manuel Nava 6, San Luis Potos\'{\i}, S.L.P. 78290, M\'exico}

\author{T.~Kl\"ahn}
\email{thomas.klaehn@googlemail.com}
\affiliation{
Institute for Theoretical Physics,
University of Wroc{\l}aw,
Max Born Pl. 9,
50-204 Wroc{\l}aw, Poland}

\begin{abstract}
We discuss the observation 
that under neutron star conditions of charge neutrality and $\beta-$equilibrium 
the contribution from the symmetry energy to the equation of state (EoS) 
follows a universal behaviour.
We call this behaviour
the conjecture of a Universal Symmetry Energy Contribution (USEC).
We find that an USEC holds provided the density dependence of the symmetry energy $E_s(n)$ follows a behaviour that limits the proton fraction $x(n)$ to values below the threshold for the direct Urca (DU) cooling process.
The absence of DU cooling in typical mass neutron stars appears to be supported 
by  the phenomenology of neutron star cooling data 
and allows to constrain the behaviour of $E_s(n)$ at  high densities.
Two classes of symmetry energy functions are investigated more in detail to elucidate the USEC.
We derive an analytic formula for the USEC to the neutron star EoS based on the result for the symmetry energy extracted from isobaric analog states of nuclei.
\end{abstract}

\pacs{04.40.Dg, 12.38.Mh, 26.60.+c, 97.60.Jd}
\maketitle

\section{Introduction}
The study of matter at extreme densities has been of great importance during the
last decades. On the one hand heavy ion collisions (HIC) at relativistic energies in
the laboratory probe densities far beyond those found in nuclei. 
On the other hand, nature provides similar conditions inside neutron stars (NS), which
can be studied by means of astronomical observations. 
The link between these two systems is the nuclear symmetry energy \cite{EPJA50}
which is an important ingredient for the physics of neutron stars where matter is highly 
asymmetric in proton and neutron numbers in contrast to the almost symmetric
systems in HIC~\cite{Tsang:2003td,Danielewicz:2003dd,Lattimer:2012xj}. 
The density dependence of the symmetry energy $E_s(n)$ determines the proton fraction 
$x(n)$ in compact stars and thus the way they cool. 
In the context of the the NS equation of state (EoS), direct Urca (DU) cooling~\cite{Lattimer:1991ib} 
plays a decisive role as we will show in this work.
The observation that the variation of $E_s(n)$ is largely compensated by the factor
$\alpha^2=(1-2x(n))^2$ for EoS that do not allow cooling of typical neutron stars by the DU process 
leads to the conjecture of a \textit{Universal Symmetry Energy Contribution (USEC)} $S^{U}(n)$ to the neutron star EoS \cite{Klahn:2006ir}. 

Two classes of symmetry energy functions are investigated in more detail to elucidate the USEC.
Both of them fulfil the constraint from an analysis using isobaric analog states (IAS) 
by Danielewicz and Lee \cite{Danielewicz:2013upa}  which revealed that  $E^*=E_s(n^*)=25.7~$MeV
at a reference density $n^*=0.105~$fm$^{-3}$
and is constrained to a rather narrow band at subsaturation densities. 
The first one follows a power law ansatz (which we refer to as "MDI type"),
$E_s(n) = E^* \cdot \left(n/n^*\right)^{\gamma}$, 
where the IAS constraint limits the admissible values of $\gamma$ to the 
wide range $1/6 \le \gamma \le 9/10$ when focussing on variations at saturation density 
$n_0=0.15~$fm$^{-3}$ only.
This range gets narrowed to $2/3 \le \gamma \le 9/10$ when also smaller variations at the lower 
limit $n=n_0/4$ for the IAS constraint are respected.
This ansatz allows for the DU process at densities below $3n_0$ which without further constraints on the stiffness of the nuclear EoS would occur also in typical mass neutron stars and thus apparently violate 
the USEC.
The second one uses a recent parametrization of the density-dependent couplings in the isovector $\rho$ 
meson channel within the generalized density functional approach to nuclear matter \cite{Typel:2014tqa} 
leading to a moderate increase of the symmetry energy at supersaturation densities; gentle enough to fulfil the DU constraint in the whole range of densities relevant for neutron star interiors and thus in perfect agreement with the USEC. 
We will denote it as "DD2-type".

This paper is organized as follows:  
In section II we recollect the basic relationships for a general equation of state of compact star matter,
i.e., degenerate nuclear matter in $\beta$-equilibrium with electrons and muons under the condition of
electric charge neutrality. 
We derive a general relationship between the nuclear symmetry energy and the proton fraction in the 
parabolic approximation for the dependence on the asymmetry $\alpha$.
In section III we consider the strong interaction part of the symmetry energy and contrast two classes
of symmetry energy models: 
A) the power-law behaviour and 
B) the density-dependent RMF model DD2~\cite{Typel:2014tqa}.
Both fulfill recent analyses of isobaric analogue states \cite{Danielewicz:2013upa}.
In section IV we show the dependence of neutron star properties on the symmetry energy,
discuss the interrelation between these two EoS and the direct Urca process,
and determine the conditions under which the symmetry energy contribution to the EoS behaves universal.
We derive an analytic formula for this universal contribution as the main result of this work.
In the final section we summarize the results and draw conclusions from this work.

\section{Condition of charge neutrality and $\beta-$ equilibrium}
Starting point is the energy per nucleon in cold neutron star matter
\bea
\label{etot}
E_{\rm tot}(n,\{x_i\}) &=& E_{\rm b}(n,x_p) + E_{\rm lep}(n,x_e,x_\mu)~,\\
E_{\rm b}(n,x_p)&=&E_0(n) + S(n,x_p) \\
E_{\rm lep}(n,x_e,x_\mu)&=&E_e(n,x_e) + E_\mu(n,x_\mu) ~,
\label{elep}
\eea
where $n=n_p+n_n$ is the total baryon density and $x_i=n_i/n$, $i=n, p, e, \mu$ 
are the fractions of protons, electrons and muons, respectively.
The dependence of the baryonic part $E_{\rm b}(n,x_p)$ on the asymmetry 
\bea
\alpha(x_p)&=&\frac{n_n-n_p}{n_n+n_p}=1-2x_p,
\eea
is used here to define the symmetry energy as
\bea
E_s(n) = \frac{1}{2}\frac{\partial^2 S(n,x_p)}{\partial \alpha^2}\bigg|_{\delta=0}~.
\eea
For the parabolic approximation this results in
\bea
\label{sym}
S(n,x_p)&=&(1-2x_p)^2 E_s(n)~.
\eea
The leptonic contribution is a sum of the Fermi gas expressions for the contributing leptons $l=e,\mu$
\bea
E_l(n,x_l)&=& \frac{1}{n}\frac{p_{F,l}^4}{4\pi^2}\left[\sqrt{1+z_l^2}\left(1+\frac{z_l^2}{2}\right)\right.
\nonumber\\
&&\left. -\frac{z_l^4}{2}{\rm Arsinh}\left(\frac{1}{z_l}\right) \right] ~,
\eea
where $z_l=m_l/p_{F,l}$. For massless leptons ($z_l\to 0$), this expression goes over to
\bea
E_l(n,x_l)\big|_{m_l=0}= \frac{1}{n}\frac{p_{F,l}^4}{4\pi^2} 
= \frac{3}{4}\left(3\pi^2 n\right)^{1/3} x_l^{4/3}~.
\eea
Under neutron star conditions charge neutrality holds,
\bea
\label{neut}
x_p=x_e+x_\mu~.
\eea
The $\beta-$ equilibrium with respect to the weak interaction processes
$n\to p+ e^- +\bar{\nu}_e $ and $p+e^- \to n+\nu_e$ (and similar for muons)
for cold neutron stars (temperature $T$ below the neutrino opacity criterion $T<T_\nu\sim 1$ MeV) 
implies
\bea
\label{beta}
\mu_n - \mu_p = \mu_e = \mu_\mu~.
\eea
The chemical potentials are defined as
\bea
\mu_i = \frac{\partial \varepsilon_i}{\partial n_i} = \frac{\partial}{\partial x_i} E_i(n,\{x_j\}) ~,~~i,j=n,p,e,\mu~,
\eea  
where $\varepsilon_i = n\ E_i(n,\{x_j\}) $ has been introduced as the partial energy density of species $i$
in the system. 
Due to the charge neutrality relation (\ref{neut}) we can eliminate the electron fraction $x_e=x_p-x_\mu$
and denote for brevity the proton fraction as $x=x_p$ in the following.
The total energy (\ref{etot}) at fixed baryon density is then a function of two variables, the proton and electron fractions, which follow from the stationarity conditions
\bea
\label{a}
\frac{\partial E_{\rm tot}}{\partial x}\bigg|_{x_\mu
} &=& 0 = -4(1-2x) E_s(n) 
+\frac{\partial E_e}{\partial x_e}\frac{\partial x_e}{\partial x}~,\\
\frac{\partial E_{\rm tot}}{\partial x_\mu}\bigg|_{x
} &=& 0 = 
\frac{\partial E_e}{\partial x_e}\frac{\partial x_e}{\partial x_\mu}+ \frac{\partial E_\mu}{\partial x_\mu}~,
\label{b}
\eea
being equivalent to the $\beta-$ equilibrium conditions (\ref{beta}). While Eq.~(\ref{b}) means
$\mu_e=\mu_\mu$, Eq.~(\ref{a}) relates the symmetry energy and the proton fraction with the electron
chemical potential
\bea
\mu_e = 4 (1-2x) E_s(n)~.
\eea
Since electrons in neutron star interiors are ultrarelativistic, 
$\mu_e=\sqrt{p_{F,e}^2 + m_e^2}\approx p_{F,e}$, and 
\bea
p_{F,e} = (3\pi^2 n_e)^{1/3} = (3\pi^2 n)^{1/3} (x-x_\mu)^{1/3}~, 
\eea
We arrive at  a system of two equations which determine the proton and muon fractions as functions of the 
baryon density once the symmetry energy $E_s(n)$ is known
\bea
\label{x}
\frac{x-x_\mu}{(1-2x)^3}&=&\frac{64 E_s^3(n)}{3\pi^2 n}~,\\
(x-x_\mu)^{2/3}-x_\mu^{2/3}&=&x_{\rm thr, \mu}^{2/3}~.
\label{xmu}
\eea
The second equation (\ref{xmu}) follows from (\ref{b}) and contains the threshold proton fraction above which muons appear in the system
\bea
\label{mu-thr}
x_{\rm thr, \mu} = \frac{m_\mu^3}{3\pi^2n}~.
\eea
For $n<n_{\rm thr,\mu}$, the muon fraction is zero and we recover the widely known relationship 
for so-called $n-p-e$ matter in $\beta-$ equilibrium
 \bea
\label{xonly}
\frac{x}{(1-2x)^3}&=&\frac{64 E_s^3(n)}{3\pi^2 n}~.
\eea
Based on these relationships we discuss now two classes of generic behaviour of the high-density symmetry energy $E_s(n)$ and their relation to the phenomenology of neutron stars.
Throughout this article we will use for $E_0(n)$ the relativistic density functional approach DD2 
with the parametrization of its  density dependent meson nucleon couplings as given in 
Ref.~\cite{Typel:2009sy}.

\section{Generic examples for $E_s(n)$ }
\label{sec:esym}

\subsection{MDI-type symmetry energy}

\begin{figure}[!htb]
\includegraphics[width=0.7\textwidth, angle=0]{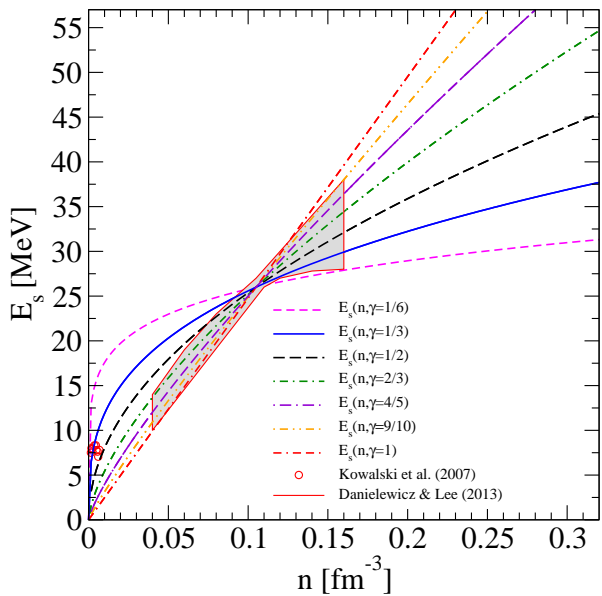}
\caption{Symmetry energy as a function of baryon density for the MDI-type ansatz (\ref{MDI}) compared to the IAS constraint \cite{Danielewicz:2013upa}.
The data points at low densities are from Ref.~\cite{Kowalski:2006ju}, for recent updates see also \cite{Hagel:2014wja}.
\label{Es-DanLee}}
\end{figure}

We base the discussion of the symmetry energy at high densities on the constraints recently obtained 
by Danielewicz and Lee \cite{Danielewicz:2013upa} by analysing isobaric analogue states 
of nuclei. 
They conclude that at a reference density $n^*=0.105$ fm$^{-3}$ the symmetry
energy is $E_s^*=E_s(n^*)=25.7$ MeV with a rather narrow error band in the density range
$0.04 < n/{\rm fm}^{-3}<0.16$, see the red area in Fig.~\ref{Es-DanLee}.

\begin{figure}[!htb]
\includegraphics[width=0.7\textwidth, angle=0]{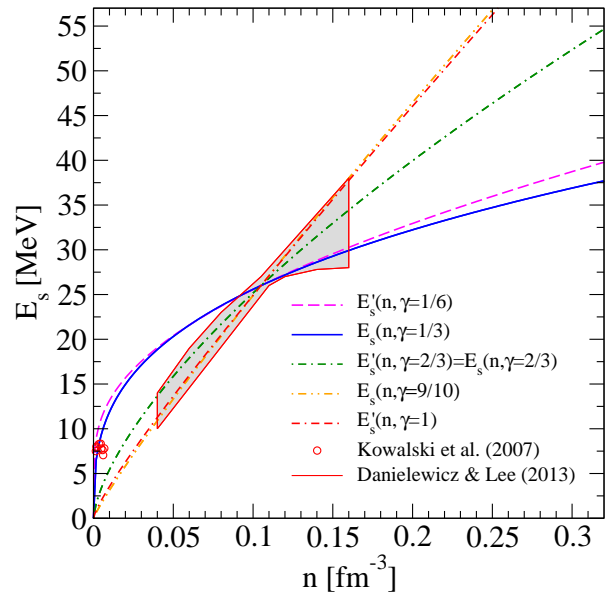}
\caption{Symmetry energy $E_s^\prime(n)$ where the kinetic part is separated from an MDI-type ansatz for the potential 
part (\ref{MDI2}) compared to MDI-type ansatz for the total symmetry energy $E_s(n)$ (\ref{MDI}). 
\label{Es-compare}}
\end{figure}

In that figure we show the symmetry energy behaviour following from the MDI-type 
\cite{Tsang:2003td} power-law ansatz 
\bea
\label{MDI}
E_s(n,\gamma) = E_s^* (n/n^*)^\gamma~,
\eea
where the exponent $\gamma$ is varied in the range $1/6 < \gamma < 1$.
One may argue that this ansatz would have to be refined by differentiating between the kinetic and potential energy contributions to the symmetry energy which would suggest the behaviour
\bea
\label{MDI2}
E_s^\prime(n,\gamma) &=& A^* (n/n^*)^{2/3}+B^* (n/n^*)^\gamma~,\\
&=&A(n/{\rm fm}^{-3})^{2/3}+B (n/{\rm fm}^{-3})^\gamma~,
\eea
with $A=42.4$ MeV and $B=16.5/(0.105)^\gamma$ MeV.
Comparing this refined ansatz with (\ref{MDI}) one recognizes that trivially for $\gamma=2/3$ 
there is no difference and that for the densities of our interest, just above the saturation density 
$n_0=0.15$ fm$^{-3}$, and a wide range of $\gamma$ values (\ref{MDI2}) can be mapped to 
(\ref{MDI}) by redefining $\gamma$ to good accuracy.
For example, $E_s^\prime(n,\gamma=1)\sim E_s(n,\gamma=9/10)$,
$E_s^\prime(n,\gamma=2/3) = E_s(n,\gamma=2/3)$  and 
$E_s^\prime(n,\gamma=1/6)\sim E_s(n,\gamma=1/3)$, see Fig.~\ref{Es-compare}.

\subsection{DD2-based symmetry energy}

\begin{figure}[!htpb]
\includegraphics[width=0.7\textwidth, angle=0]{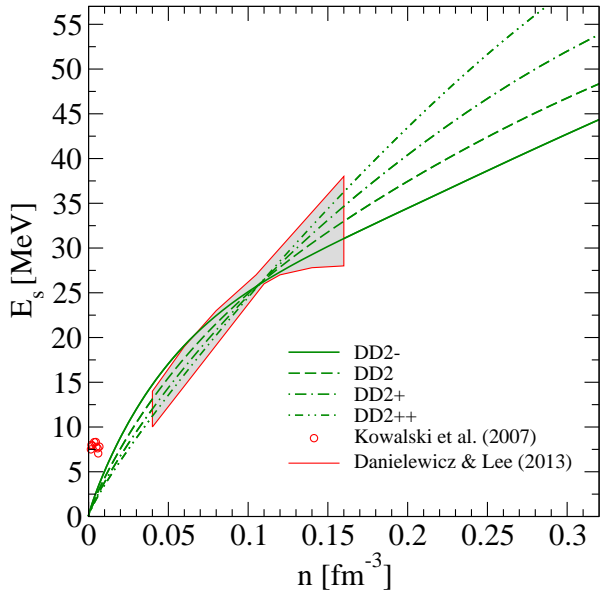}
\caption{Same as Fig.~\ref{Es-DanLee} but for the symmetry energy parametrization of the 
generalized density functional approach \cite{Typel:2014tqa}.
\label{UEsMuons}}
\end{figure}

The second example for a generic class of functions describing the high-density behaviour of the nuclear symmetry energy is based on a generalized density functional approach \cite{Typel:2009sy}.
Recently, variations of its symmetry energy relevant couplings in the $\rho-$ meson channel have been introduced in the context of discussing the neutron skin thickness of heavy nuclei \cite{Typel:2014tqa}. 
Subsequently, it was also used in exploring symmetry energy effects in simulations of core-collapse supernovae \cite{Fischer:2013eka}.
The density dependence of the symmetry energies introduced and labelled in Ref.~\cite{Typel:2014tqa}
are shown in Fig.~\ref{UEsMuons}. 

We note that the behaviour of the DD2-based symmetry energies bears striking similarities with 
other microscopic approaches based on realistic NN forces like the celebrated APR EoS \cite{Akmal:1998cf}, or the recent Brueckner-Hartree-Fock calculation employing the Argonne V18 NN potential 
supplemented with three-nucleon forces \cite{Logoteta:2015voa}.

\subsection{Derived quantities: $J$, $L$ and $K_{\rm sym}$}

\begin{figure}[!htb]
\includegraphics[width=0.7\textwidth, angle=0]{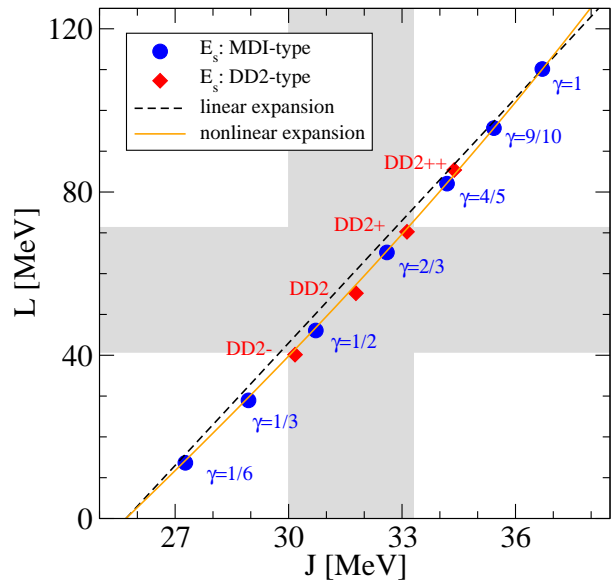}
\caption{Relation between L and J parameters for the symmetry energy models discussed in the text.
The dashed line is an analytic approximation derived 
from a Taylor expansion of the symmetry energy (\ref{L-S}) using the constraint derived by 
Danielewicz and Lee \cite{Danielewicz:2013upa} and dropping the curvature term whereas the solid line 
corresponds to Eq.~(\ref{L-S2}) which includes the expression (\ref{Ksym2}) for $K_{\rm sym}$. 
For shaded regions, see text.
\label{fig:S-L}}
\end{figure}

In order to characterize the above models for the symmetry energy we compute its standard 
parameters, the value $J$, the slope $L$ and the curvature $K_{\rm sym}$ of the symmetry energy,
defined at nuclear saturation density as 
\bea
J &=& E_s(n_0)~,\\
L &=& 3 n \frac{\partial E_s(n)}{\partial n}\bigg|_{n=n_0}~,\\
K_{\rm sym} &=& 9 n^2 \frac{\partial^2 E_s(n)}{\partial n^2}\bigg|_{n=n_0}~.
\eea 
In Table~\ref{tab:parameters} we list the corresponding values obtained for the parametrizations of the 
two types of symmetry energy models used in this work and give also experimental values with their 
references.
\begin{table}[!t]
\label{tab:parameters}
\caption{Symmetry energy parameters derived from experiments$^*$ (upper part) compared to those for the MDI-type models (middle part, different $\gamma$-values) and for the DD2-type models (lower part).}
\label{SLKparameters}
\begin{tabular}{lccc}
\hline \hline
 &$J$ &$L$ & $K_{\rm sym}$\\
model & [MeV] & [MeV] & [MeV] \\
\hline
\cite{Danielewicz:2013upa}&		$33\pm5 $	& $52.5\pm17.5$	& $-235 \pm100$ \\
&&&$(-106.18 \pm 58.28)$\\
\cite{Roca-Maza:2015eza,Ban:2010wx}&	$31\pm 2$	& $43\pm 6$	& $-292 \pm 100$\\
&&&$(-137.81 \pm 19.98)$\\
\cite{Carbone:2010az}&			$32.2\pm 1.3$	& $64.8\pm 15.7$& $-161 \pm 100 $\\
&&&$(-65.22 \pm 52.28)$\\
\cite{Tsang:2008fd}&			$32\pm 2$	& $70 \pm 30$	& $-130 \pm 100$ \\
&&&$(-47.90 \pm 99.90)$\\
\cite{Chen:2010qx}&			$30\pm 4$	&$58 \pm 18$	& $-202 \pm 100 $\\
&&&$(-87.86 \pm 59.94)$\\
\hline
$\gamma=$1/6&	27.27& 13.64& -34.09\\
$\gamma=$1/3&	28.94& 28.94& -57.89\\
$\gamma=$1/2&	30.72& 46.08& -69.11\\
$\gamma=$2/3&	32.60& 65.20& -65.20\\
$\gamma=$8/10&	34.19& 82.05& -49.23\\
$\gamma=$9/10&	35.43& 95.66& -28.67\\
$\gamma=$1&	36.71& 110.14& 0\\
\hline
DD2-&	30.17& 40.14& -54.41\\
DD2&	31.78& 55.19& -93.33\\
DD2+&	33.13& 70.25& -93.12\\
DD2++&	34.38& 85.40& -64.59\\
\hline \hline
\end{tabular}
\\[2mm]
\small{$^{*} K_{\rm sym}$ is derived from the measured values of $L$ and $K_{\rm \tau}$ as 
$K_{\rm sym}=K_{\rm \tau}+6L$~\cite{Baran:2001pz} or as
$K_{\rm sym} =3.33 L - 281$ MeV~\cite{Ducoin:2011fy} (in parentheses).}
\end{table}
Using our knowledge of the symmetry energy at the Danielewicz-Lee reference point $n^*$,
we derive a relationship between the $J$ and $L$ parameters from the Taylor expansion of the symmetry energy
\bea
E_s(n^*) &=& E_s(n_0) + (n^* - n_0)\frac{\partial E_s(n)}{\partial n}\bigg|_{n=n_0}\nonumber\\
&&+\frac{(n^*-n_0)^2}{2}\frac{\partial^2 E_s(n)}{\partial n^2}\bigg|_{n=n_0}+\dots~,\nonumber\\
&=& J + \frac{n^*-n_0}{3n_0} L+\frac{(n^*-n_0)^2}{18n_0^2}K_{\rm sym}+\dots~. \nonumber\\
\eea
Using the fact that $n_0-n^*= 3n_0/10$ this results in 
\bea
\label{L-S}
L=10(J-E_s^*)+\frac{1}{20}K_{\rm sym}~.
\eea
Neglecting the curvature term results in a linear approximation which is in excellent agreement with 
the numerical results for the $L-J$ relation of both models for the symmetry energy as shown in 
Fig.~\ref{fig:S-L}, see also \cite{Horowitz:2014bja}.

We estimate the curvature term analytically for the MDI-type symmetry energy (\ref{MDI}) where
\bea
\label{Ksym}
K_{\rm sym}=9\gamma(\gamma-1)J~.
\eea
Using the relationship
\bea
\label{S-E*}
J=E_s^*(n_0/n^*)^\gamma \simeq E_s^*(1+3\gamma/7)~,
\eea
we eliminate $\gamma$ from (\ref{Ksym}) and obtain
\bea
\label{Ksym2}
K_{\rm sym}&=&\frac{7 J}{{E_s^*}^2}(J-E_s^*)(7J-10E_s^*)~,\nonumber\\
&=&\frac{49 J}{{E_s^*}^2}\left(J-\frac{17E_s^*}{14}\right)^2 -\frac{9J}{4}~.
\eea
With Eq.~(\ref{L-S}) we obtain the nonlinear $L-J$ relationship
\bea
\label{L-S2}
L=\frac{791}{80}\left(J-\frac{800E_s^*}{791}\right)+\frac{49J}{20{E_s^*}^2}\left(J-\frac{17E_s^*}{14} \right)^2~,
\eea
which is shown together with the linear approximation to Eq.~(\ref{L-S}) in Fig.~\ref{fig:S-L}.
Note that the linear approximation to (\ref{L-S}) has been reported before in \cite{Horowitz:2014bja}, 
while the nonlinear $L-J$ relation (\ref{L-S2}) is a new result of the present work. 

The shaded regions shown in Fig.~\ref{fig:S-L} follow from the requirements that (i) the model for the symmetry energy $E_s(n)$ shall not leave the region of the Danielewicz-Lee constraint shown in 
Figs.~\ref{Es-DanLee} - \ref{UEsMuons} and (ii) that central densities in typical neutron stars shall not exceed the threshold density for the rapid direct Urca cooling process, see next section. 

\section{Results for neutron star properties}

\subsection{Neutron star masses and radii vs. $E_s(n)$}

\begin{figure}[!htb]
\includegraphics[width=0.7\textwidth, angle=0]{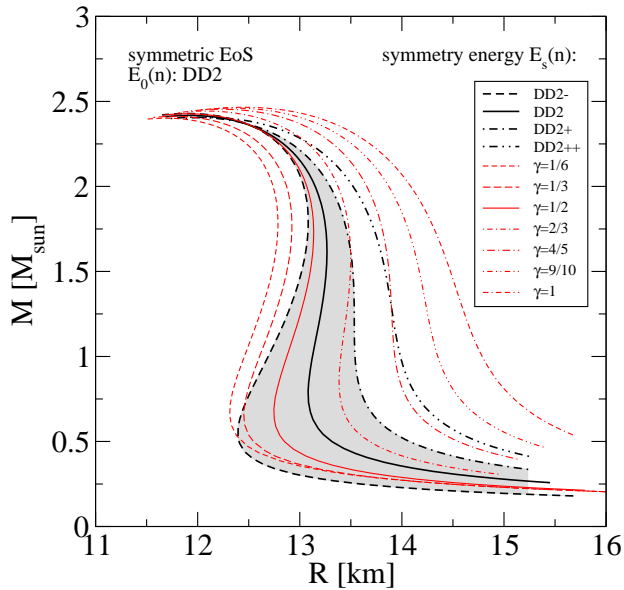}
\caption{(Color online) Neutron star mass - radius relation for the different symmetry energies discussed in the text. The symmetric EoS is taken to be DD2. The grey region gives an estimate of the limited variation 
for the $M-R$ sequence that would be induced by a variation of the symmetry energy in accordance with the Danielewicz-Lee constraint and the direct Urca cooling constraint. For details, see text.
\label{m-r}}
\end{figure}

Given the equation of state of neutron star matter, the structure and global properties of compact stars are obtained from solving the Tolman-Oppenheimer-Volkoff (TOV) equations 
\cite{Tolman:1939jz,Oppenheimer:1939ne}
\bea
\frac{dP( r)}{dr}&=& - \frac{G M( r)\varepsilon( r)}{r^2}\frac{\left(1+\frac{P( r)}{\varepsilon( r)}\right)
\left(1+ \frac{4\pi r^3 P( r)}{M( r)}\right)}{\left(1-\frac{2GM( r)}{r}\right)},\nonumber\\
\\
\frac{dM( r)}{dr}&=& 4\pi r^2 \varepsilon( r),\\
\frac{d N_B( r)}{dr}&=& 4\pi r^2 \left(1-\frac{2GM( r)}{r}\right)^{-1/2}n( r)~.
\eea
Starting with a central energy density $\varepsilon_c=\varepsilon(r=0)$ and pressure $P_c=P( r=0)$ as boundary value at $r=0$, these equations are 
integrated out to the distance $r=R$ where the pressure vanishes $P( r=R)=0$, defining the radius $R$,
the mass $M=M( R)$ and the baryon number $N_B=N_B( R)$ 
of the star.
Varying the central energy density one obtains a sequence of star configurations for a given EoS, which then uniquely corresponds to a mass-radius curve $M( R)$ \cite{Lindblom:1992}, 
as shown in Fig.~\ref{m-r} for the classes of EoS under investigation in this work.
The EoS for symmetric matter $E_0(n)$ is given by the DD2 EoS and for $E_s(n)$ we use either the MDI-type or the DD2-type functions.
The grey shaded region in Fig.~\ref{m-r} is bordered by the $M-R$ sequences for the DD2- and the DD2+ symmetry energies which mark the corner points for the grey band highlighted in the $L-J$ plane of 
Fig.~\ref{fig:S-L}. 
As for the description of the neutron star crust which plays an important role for the determination of the radii, we have taken the SLy EoS composed of different parts, including the well established results from \cite{hp} and \cite{dh}.
The crust table, together with this crust EoS full description, can be found in \cite{ioffe}.
Further detailed discussion of the relation between symmetry energy and neutron star radius can be found, e.g., in \cite{AlvarezCastillo:2012rf}. 

\begin{figure}[!htbp]
\includegraphics[width=0.47\textwidth, angle=0]{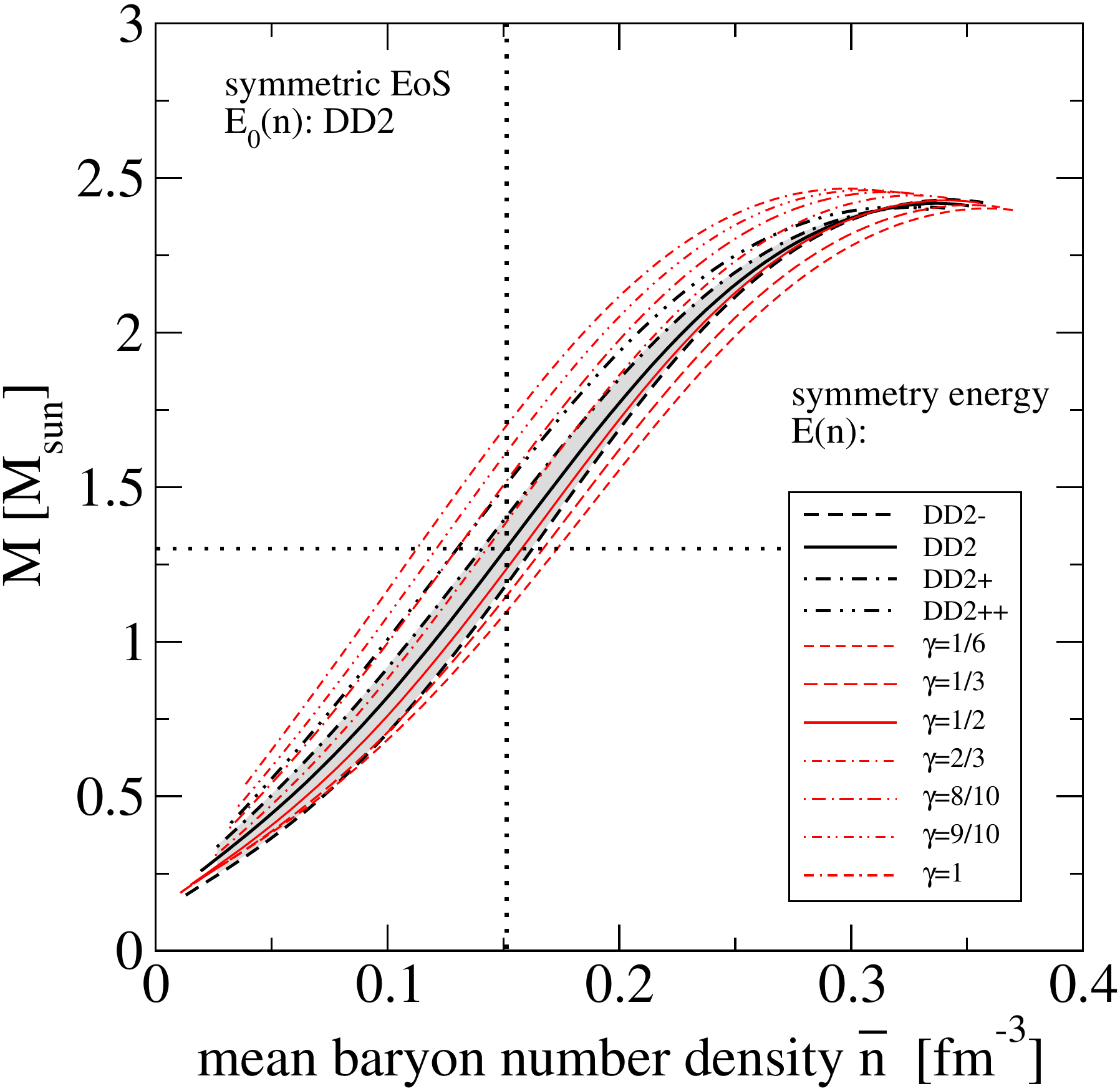}
\caption{Correlation of mass and mean density of the neutron star for the different symmetry energies discussed in the text. The symmetric EoS is taken to be DD2.
Note that for the DD2 EoS the neutron star with $M=1.3~M_\odot$ has a mean baryon number density 
equal to the saturation density, $\bar{n}=n_0=0.15$ fm$^{-3}$.
\label{m-n}}
\end{figure}

In order to quantify the relationship between neutron star radii and the nuclear symmetry energy, we first
introduce the notion of a mean baryon density for a neutron star configuration characterized by its mass 
and baryon density profile, 
\bea
\bar{n} = \frac{N_B}{V}=\frac{\int_0^R dr r^2 \left(1-\frac{2GM( r)}{r}\right)^{-1/2}n( r)}{\int_0^R dr r^2 \left(1-\frac{2GM( r)}{r}\right)^{-1/2}}~.
\eea
In Fig.~\ref{m-n} we show the gravitational mass as a function of the mean baryon number density for the same star sequences as in Fig.~\ref{m-r}.
It is remarkable that despite the large spread in radii for stars of the typical binary radio pulsar mass 
(mean $M_{\rm BRP}\sim 1.4~M_\odot$) their variation in mean baryon densities is rather well centered around the saturation density. 
Thus if we describe a star by the DD2 EoS and ask for which gravitational mass the mean density equals the saturation density $n_0=0.15$ fm$^{-3}$ of that model, we find a mass of $1.3~M_\odot$, see Fig.~\ref{m-n}.

\subsection{Direct Urca process constraint}

With the two generic classes of symmetry energy behaviour introduced in the previous section we can now solve Eqs.~(\ref{x}) and (\ref{xmu}) for the proton and muon fractions under neutron star conditions. 
The proton fraction plays an important role in the neutron star phenomenology as it determines whether the fastest neutrino cooling process, the direct Urca (DU) process $n\to p + e + \bar{\nu}_e$, can occur or not \cite{Lattimer:1991ib}.  
If the central density of a neutron star exceeds the critical value and triggers the DU process this causes a dramatic drop of the core temperature due to rapid energy loss by 
neutrino emission. After the typical transport timescale of about 100 years the cooling wave reaches the surface of the star and the photon luminosity drops rapidly making the star practically invisible.
This process can therefore not be operative in typical neutron stars as we do observe cooling neutron stars much older than 1000 years with surface temperatures that are not compatible with the DU cooling scenario.
For a detailed discussion of the DU process constraint see, e.g., Refs.~\cite{Klahn:2006ir,Blaschke:2006gd}.
\begin{figure}[!htpb]
\includegraphics[width=0.7\textwidth, angle=0]{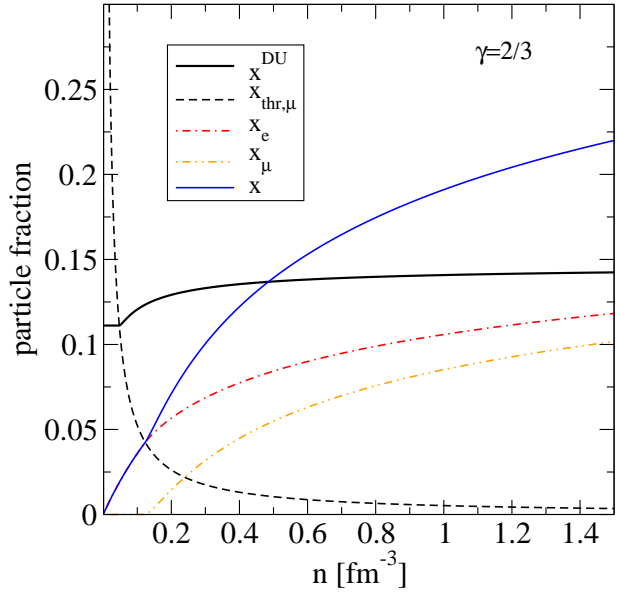}
\caption{Particle fractions in neutron star matter as a function of the baryon density for the MDI-type symmetry energy parametrization with $\gamma=2/3$. Also shown are the DU threshold (solid line) and the muon threshold (dashed line) for the proton fraction. 
\label{x-gamma23}}
\end{figure}
\begin{figure}[!thpb]
\includegraphics[width=0.7\textwidth, angle=0]{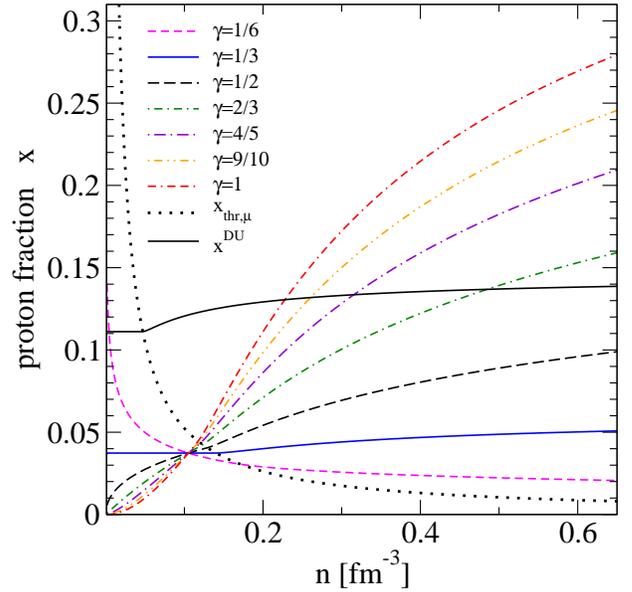}
\includegraphics[width=0.7\textwidth, angle=0]{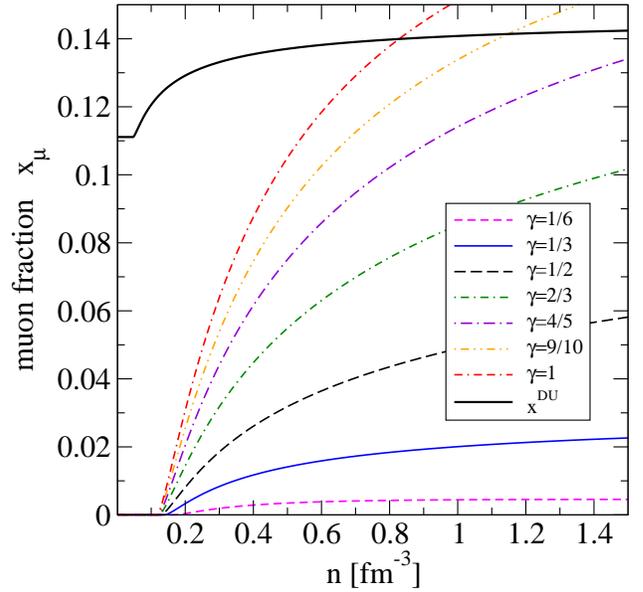}
\caption{Upper panel: proton fraction as a function of baryon density for the MDI-type ansatz (\ref{MDI})
compared to the proton fraction threshold of the direct Urca process (thin solid line). 
Also shown is the threshold proton fraction for the appearance of muons in neutron star matter 
(thin dashed line). 
Lower panel: muon fraction as a function of the baryon density for the MDI-type ansatz (\ref{MDI}).
\label{x-gamma}}
\end{figure}
\begin{figure}[!htb]
\includegraphics[width=0.7\textwidth, angle=0]{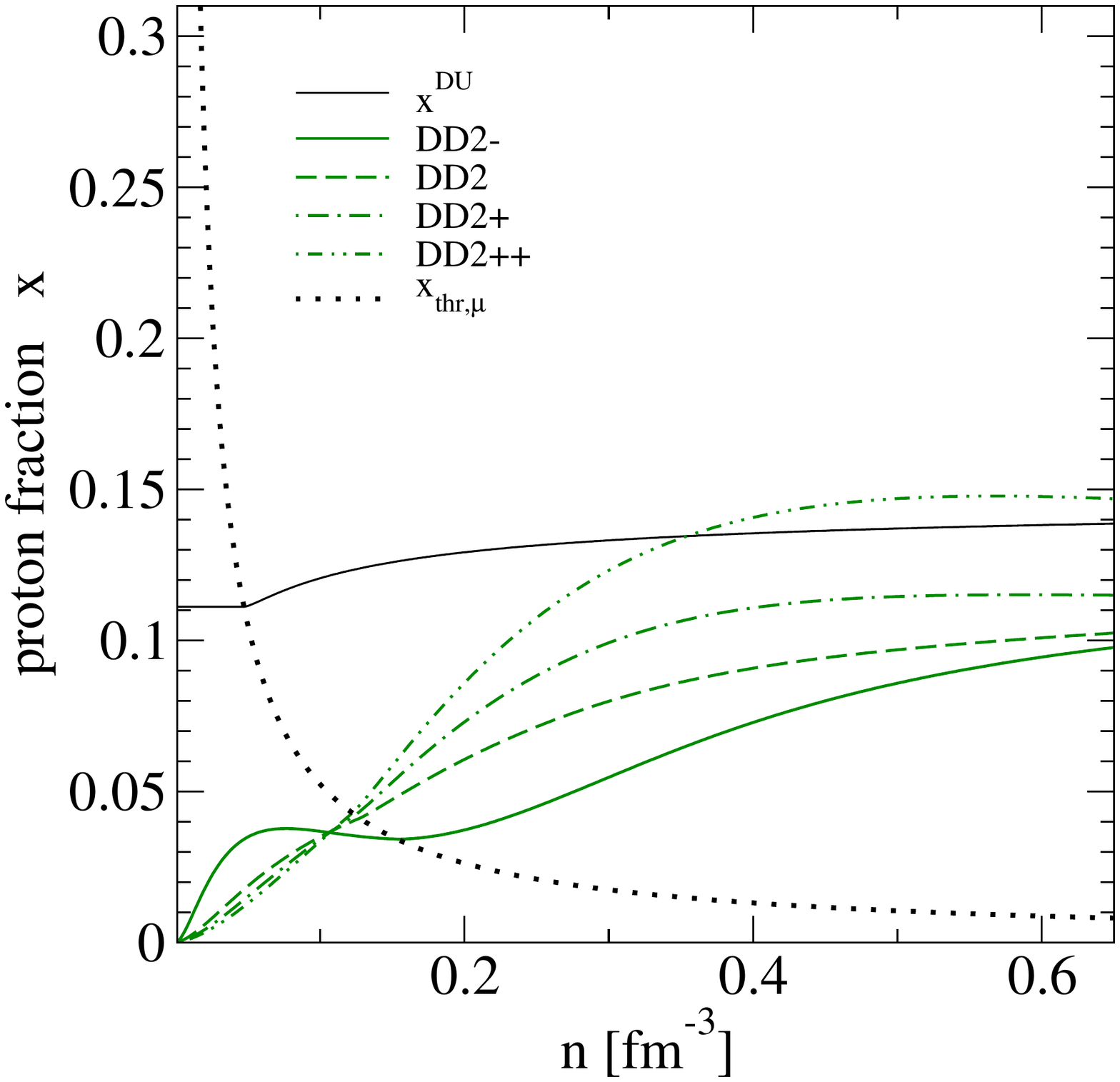}
\includegraphics[width=0.7\textwidth, angle=0]{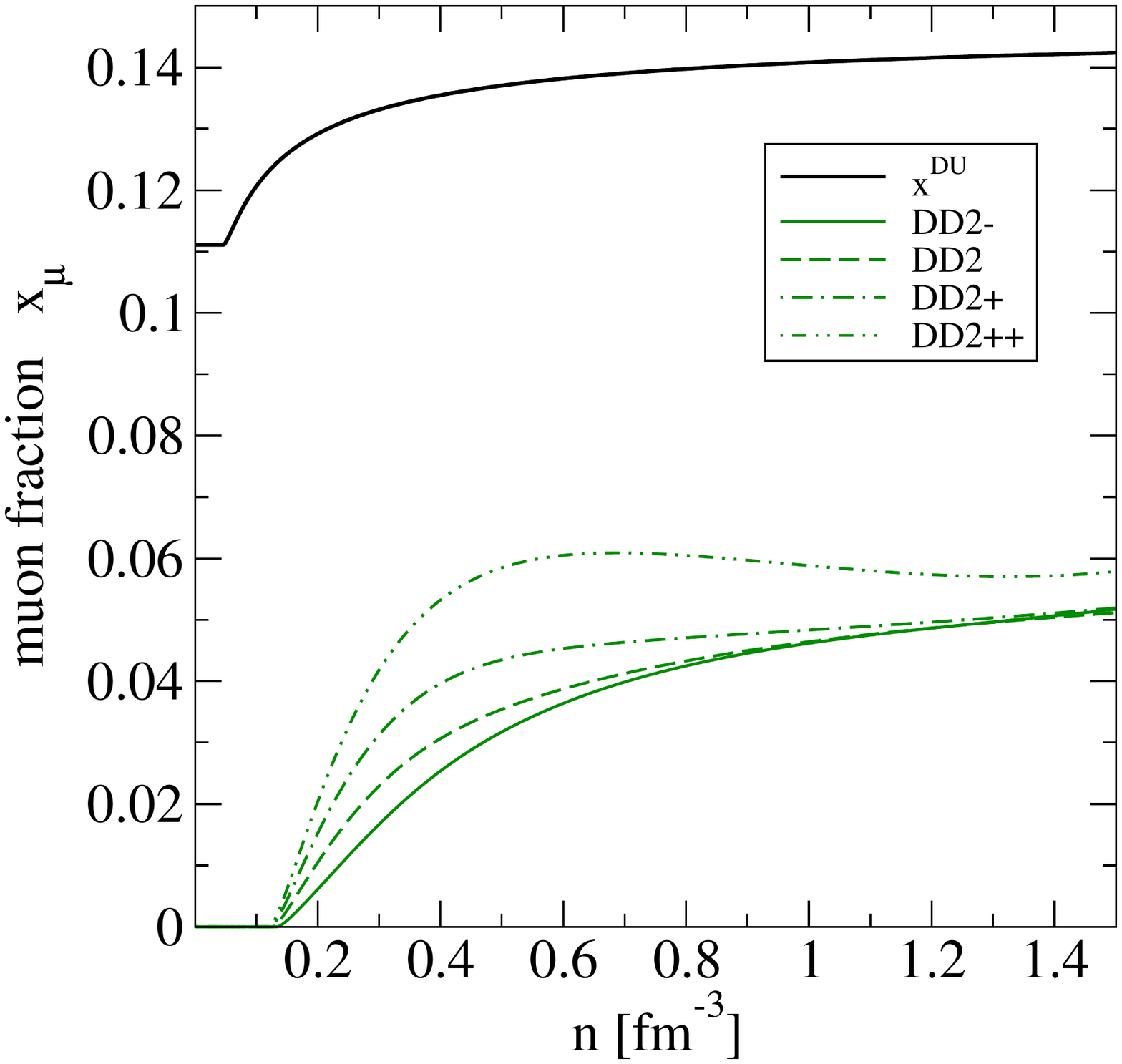}
\caption{Same as Fig.~\ref{x-gamma} but for the symmetry energy parametrization of the 
generalized density functional approach \cite{Typel:2014tqa}.
\label{x-DD2}}
\end{figure}

The key relation for deriving the DU threshold condition is the triangle inequality for the Fermi momenta of neutron, proton and electron involved in the process (neutrino momenta are small in comparison to these and can be safely neglected) leading to the condition
\bea
n_n^{1/3} < n_p^{1/3} + n_e^{1/3}~,
\eea
which can be formulated in terms of proton and muon fractions as $(1-x)^{1/3}<x^{1/3}+ (x-x_\mu)^{1/3}$,
being equivalent to  
\bea
x>\frac{1}{1+[1+(1-x_\mu/x)^{1/3}]^3}~.
\eea
For densities below the muon threshold one easily recovers the classical result for the DU
threshold without muons $x^{\rm DU}=1/9=11.1\%$ \cite{Lattimer:1991ib}.
Inserting the density dependent muon fraction $x_\mu(n)$ one obtains the density dependence of the 
DU threshold shown in Figs.~\ref{x-gamma23}, \ref{x-gamma} and \ref{x-DD2}. 
The density dependence of the particle fractions and the DU- and muon thresholds for the proton fraction are shown in Fig.~\ref{x-gamma23} for the MDI-type parametrization with $\gamma=2/3$.

For $\gamma=1/3$ the proton fraction without muons (i.e., for $n^*<n_{\rm thr,\mu}$) is density independent. Its value can be found from solving Eq.~(\ref{xonly}) with $E_s(n)=E_s^*(n/n^*)^{1/3}$ and amounts to $x=0.0363$. 
As we see from Fig.~\ref{x-gamma} (upper panel), the onset of muons does practically not change this.
The muon fraction reaches $x_\mu=x/2$ for asymptotically large $n$, see Eq.~(\ref{xmu}).

When comparing the behaviour of the proton fractions for the MDI-type models in Fig.~\ref{x-gamma} 
with those of the DD2-based models in Fig.~\ref{x-DD2} one observes a striking difference.
While for the former the DU constraint is violated for all $\gamma \ge 2/3$ already for densities below 
$\sim 3 n_0$, the DU process is excluded for all DD2-based models with exception of the model DD2++  which has a proton fraction that touches the DU constraint for densities above $\sim 2n_0$.
\begin{table}[htbp!]
\centering
\caption{Threshold densities $n_{\rm DU}$ for the direct Urca cooling process compared to central densities $n_c$ 
of compact stars with masses $M/M_\odot=1.25, 1.40, 1.60, 1.80, 2.00$, respectively.
The bold numbers indicate central densities exceeding the corresponding direct Urca threshold so that 
in these stars this fast cooling process would operate.}
\label{tab:DUdensities}
\begin{tabular}{l|c|ccccc}
\hline \hline
$E_{s}$&	$n_{\rm DU}$[fm$^{-3}$] &	& &		$n_c$[fm$^{-3}$]& & \\		
&		&1.25&	1.40&	1.60&	1.80&	2.00\\
\hline
$\gamma=1/6$	&-&	0.357&	0.379&	0.412&	0.452&	0.504\\
$\gamma=1/3$	&-&	0.346&	0.369&	0.402&	0.440&	0.492\\
$\gamma=1/2$	&-&	0.334&	0.356&	0.388&	0.426&	0.476\\
$\gamma=2/3$	&0.485&	0.318&	0.342&	0.374&	0.412&	0.458\\
$\gamma=4/5$	&0.315&	0.306&	{\bf 0.330}&	{\bf 0.361}&	{\bf 0.399}&	{\bf 0.446}\\
$\gamma=9/10$	&0.260&	{\bf 0.295}&	{\bf 0.319}&	{\bf 0.352}&	{\bf 0.390}&	{\bf 0.437}\\
$\gamma=1$&	0.228&	{\bf 0.288}&	{\bf 0.311}&	{\bf 0.344}&	{\bf 0.382}&	{\bf 0.429}\\
\hline
DD2-	&-&	0.331&	0.352&	0.385&	0.423&	0.472\\
DD2	&-&	0.331&	0.354&	0.387&	0.426&	0.478\\
DD2+	&-&	0.325&	0.349&	0.384&	0.425&	0.479\\
DD2++	&0.354&	0.314&	0.339&	{\bf 0.375}&	{\bf 0.416}&	{\bf 0.469}\\
\hline \hline
\end{tabular}
\end{table}

The question arises for a comparison with the central densities of neutron stars depending on their mass.
We show these densities in Tab.~\ref{tab:DUdensities} for five neutron star masses: $1.25,~1.42,~1.60,~1.8,~2.0~M_\odot$ where the DD2 EoS which was employed for $E_0(n)$
and the results for the symmetry energy models used in this work are given.
For those models where there is no entry for $n_{\rm DU}$ in the table, the DU threshold is never reached so that the constraint can not be violated.
For the $\gamma=2/3$ model the threshold density exists, but it is high enough that the central densities even in the most massive stars of $2.0~M_\odot$ do not exceed it.
This statement, however, has to be taken with the caveat that we have used here the rather stiff $E_0(n)$ 
of the DD2 model.
For the $\gamma=9/10$ and $\gamma=1$ models the opposite conclusion applies: the DU constraint is violated in all cases, even for the very lightest stars.
The entries given by bold numbers denote a violation of the DU constraint.
The DD2++ model case is very interesting: here the DU constraint is fulfilled for typical neutron stars with masses smaller than $\sim 1.5~M_\odot$ and violated for the more massive ones. 
This border, however, could be lifted completely once we adopt a further stiffening of $E_0(n)$, e.g., by a moderate excluded volume modification. 
Then also the most massive stars would not exhibit DU cooling \cite{Alvarez-Castillo:2016}. 

\subsection{Universal symmetry energy contribution}

With the models for the density dependence of the symmetry energy $E_s(n)$ discussed in sect.~\ref{sec:esym} and the density dependent particle fractions 
under neutron star conditions we turn now to the discussion of the symmetry energy contribution (\ref{sym}) to the neutron star equation of state (\ref{etot}) and its comparison with the lepton contribution (\ref{elep}).
\begin{figure}[!htb]
\includegraphics[width=0.7\textwidth, angle=0]{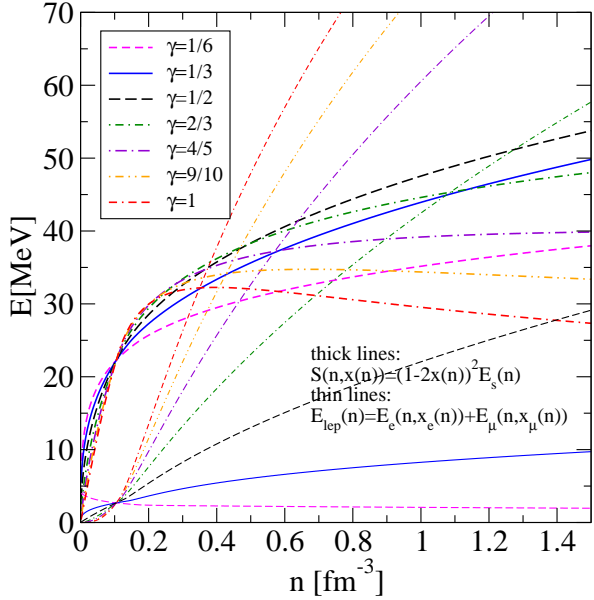}
\caption{Nuclear and leptonic energies per nucleon under neutron star conditions for the MDI-type ansatz (\ref{MDI}).
\label{E-gamma}}
\end{figure}
\begin{figure}[!htb]
\includegraphics[width=0.7\textwidth, angle=0]{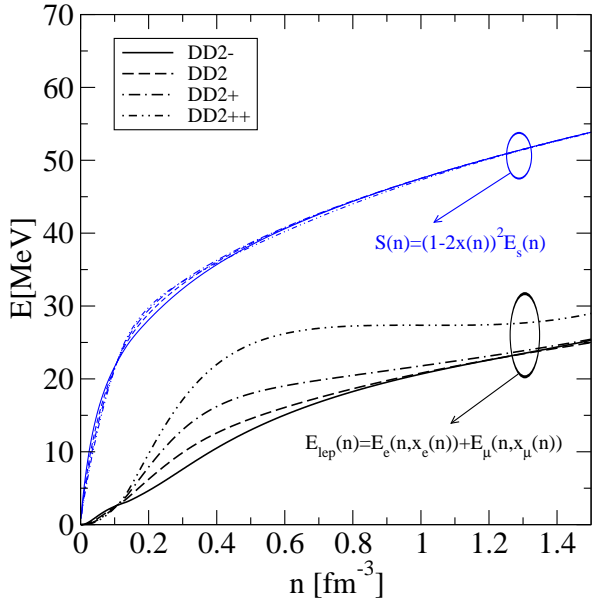}
\caption{Same as Fig.~\ref{E-gamma} but for the symmetry energy parametrization of the 
generalized density functional approach \cite{Typel:2014tqa}.
\label{E-DD2}}
\end{figure}

In Fig.~\ref{E-gamma} we show the results for the MDI-type models in the range $1/6 < \gamma < 1$.
We observe that the leptonic contribution to the EoS while being fixed at the reference density $n*$ is
otherwise wildly varying and strongly density dependent. 
The models with $\gamma \ge 2/3$ interfere with the 
symmetry energy contribution for densities below $\sim 6 n_0$ and exceed it in this range.
One may characterize the situation for these parameter values of the MDI-type model by saying that the 
neutron star behaves like a white dwarf: variations of the EoS in terms of isospin asymmetry are dominated by the lepton component!  

In the case of the DD2-based EoS shown in Fig.~\ref{E-DD2} the situation is quite different. Here the variation of the lepton contribution stays bounded and always at about half the symmetry energy contribution or even below that. The variation of the leptonic contribution which is most pronounced 
between $1-5~n_0$ is almost invisible in the symmetry energy contribution.
Therefore, we claim that for symmetry energy models which do not violate the DU constraint as the DD2-based one, 
{\bf the symmetry energy contribution to the neutron star EoS behaves universal!}
(Note that the behaviour of the symmetry energy for the DD2++ model with its gentle violation of the direct Urca constraint above $\sim 2~n_0$ marks the border of this universal behaviour.)

We would like to give an analytic understanding of this USEC. We start with the reference density 
$n^*=0.105~$fm$^{-3}$ of the Danielewicz-Lee analysis \cite{Danielewicz:2013upa} for which the symmetry energy $E_s(n^*)=25.7~$MeV is known.
We insert these values on the r.h.s. of Eq.~(\ref{xonly}) and solve the resulting cubic equation for the
proton fraction $x$ for which we obtain 
\bea
x^*=x(n^*)=0.0363~.
\eea
The muon fraction at this point is still zero \cite{footnote}.
Inserting $x^*$ into (\ref{sym}) we obtain 
\bea
\label{S*}
S^* &=& S(n^*,x^*)=(1-2x^*)^2 E_s^*\nonumber \\
 &=& \left(\frac{3\pi^2 n^*}{64} \right)^{1/3} (1-2x^*) {x^*}^{1/3}\\
 &=& 22.04~{\rm MeV}~.
\label{S*-value}
\eea
Just above the reference density $n^*$ muons appear in the system and have to be included in the analysis.
Inserting (\ref{x}) into (\ref{sym}) we eliminate the symmetry energy in favor of the proton and muon 
fractions
\begin{equation}
\label{S}
S(n,x,x_\mu) = \left(\frac{3\pi^2 n}{64} \right)^{1/3} (x-x_\mu)^{1/3} (1-2x)~.
\end{equation} 

\begin{figure}[!htb]
\includegraphics[width=0.47\textwidth, angle=0]{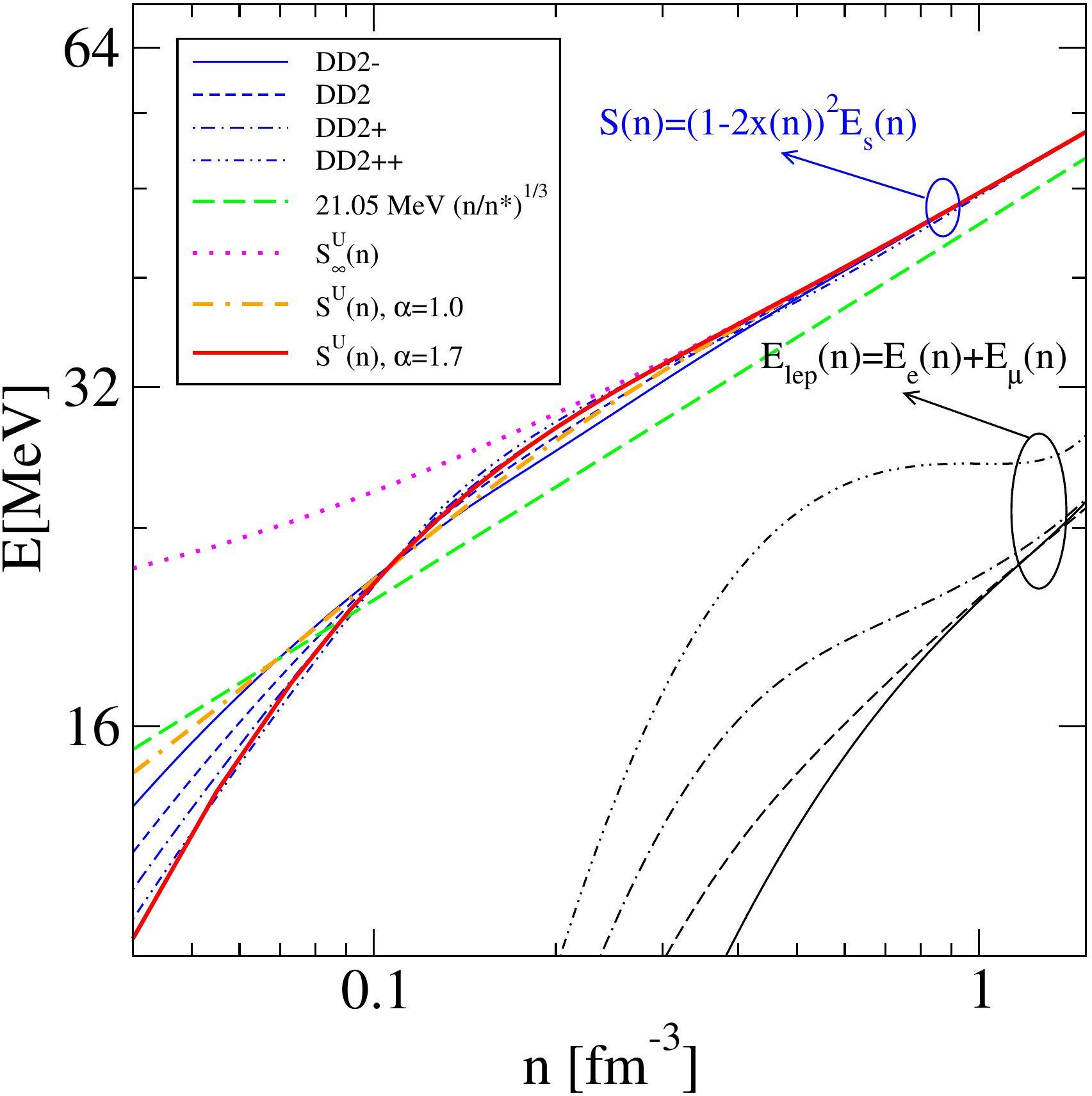}
\caption{(Color online) 
Same as Fig.~\ref{E-DD2} but enlarged in double logarithmic scale to compare the quality of 
the asymptotic solution $S_\infty^U(n)$ of Eq.~(\ref{SU}) at high densities (magenta dotted line) 
and two extrapolations $S^U(n)$  of it through the exact value $S^*$ at the reference density $n^*$ 
for the universal symmetry energy contribution to the neutron star EoS with results for the corresponding quantity from different parametrizations of the relativistic density functional approach \cite{Typel:2014tqa}.
\label{E-DD2_large}}
\end{figure}

Next we discuss the situation at asymptotically large baryon densities ($n\to \infty$), where according to  
Eq.~(\ref{xmu}) holds  $x_\mu(n\to \infty)=x_\mu^\infty =x^\infty/2$. This results in 
\begin{equation}
\label{Sinf}
S(n\to \infty,x,x_\mu=x/2)= \left(\frac{3\pi^2 n}{64} \right)^{1/3} \left(\frac{x}{2}\right)^{1/3} (1-2x)~.
\end{equation} 
The statement of universality of this contribution would mean that it is independent of the variation of the proton fraction 
\bea
\frac{\partial S(n,x,x_\mu=x/2)}{\partial x} = 0~,
\eea
which entails $x(n\to \infty)=x^\infty=1/8$ and thus $x_\mu(n\to \infty)=x_\mu^\infty = 1/16$.
Inserting these values in (\ref{Sinf}) gives 
\begin{equation}
\label{S1}
S(n,x=1/8,x_\mu=1/16) = 21.05~{\rm MeV} (n/n^*)^{1/3},
\end{equation} 
with a value at $n=n^*$ that is rather close to the exact one (\ref{S*-value}) and a quite astonishing overall agreement with the numerical solution, see Fig.~\ref{E-DD2_large}. 
Although the curve follows the trend of the USEC quite well up to the highest densities relevant for neutron stars, there is an overall deviation of about $2$ MeV.  

In order to improve the situation and to arrive at an analytical solution for the USEC to the neutron star EoS 
we follow the observation from Fig.~\ref{E-gamma} and Fig.~\ref{E-DD2} that the muon fraction stays overall closely below the massless limit case of $x_\mu=x/2$.
Let us discuss first the deviation from the asymptotic solution for $n\to \infty$.
For this purpose we introduce the deviation $\delta(x)$ from the asymptotic solution as 
$x_\mu/x=1/2-\delta(x)$ and insert this into Eq.~(\ref{xmu}) to get
\bea
\left(\frac{1}{2}+\delta(x) \right)^{2/3}-\left(\frac{1}{2} -\delta(x) \right)^{2/3} 
= \left(\frac{x^*}{x} \right)^{2/3}~. 
\eea
Expanding the l.h.s. to lowest order in $\delta(x)$, we obtain the relationship
\begin{equation}
\label{delta}
\delta(x) = \frac{3}{8} \left(\frac{2x^*}{x} \right)^{2/3}~.
\end{equation}
Inserting this asymptotic behaviour for the ratio $x_\mu/x$ in Eq.~(\ref{S}) we obtain
\bea
\label{S-n}
S(n,x) &=& \left(\frac{3\pi^2 n}{64} \right)^{1/3} 
\left(\frac{x}{2}\right)^{1/3}(1-2x)\nonumber\\
&&\times \left[1+\frac{3}{4}\left(\frac{2x_{\rm thr,\mu}}{x}\right)^{2/3}\right]^{1/3} ~.
\eea
Applying the argument of universality, i.e. vanishing variation $\partial S(n,x)/\partial x=0$, we find the relation
\bea
\label{eq:x}
0=1-8x + \left(\frac{1}{4}-5x\right)\left(\frac{2x_{\rm thr,\mu}}{x}\right)^{2/3}~,
\eea
which now yields a density dependent solution for the proton fraction. In the limit $x_{\rm thr,\mu}=0$,
i.e. for massless muons or for infinite density, the solution $x=1/8$ is recovered.
As we are looking for the density dependence $x(n)$ in the vicinity of this asymptotic solution, we 
introduce the small auxiliary quantity $\varepsilon=1/8 - x$ which after insertion in (\ref{eq:x}) fulfils the equation 
\bea
\label{eq:eps}
0=(1-8\varepsilon)^{2/3}16\varepsilon -(3-40\varepsilon)(2x_{\rm thr,\mu})^{2/3}.
\eea 
Since $\varepsilon\ll 1$, the replacement $(1-8\varepsilon)^{2/3}\to (1-16\varepsilon/3)$ is in order,.
This reduces (\ref{eq:eps}) to the quadratic equation
\bea
\varepsilon^2 - \varepsilon \left(\frac{3}{16} + \frac{15}{32} (2x_{\rm thr,\mu})^{2/3}\right)
+ \frac{9}{256} (2x_{\rm thr,\mu})^{2/3}=0~.
\nonumber\\
\eea 
Since $x_{\rm thr,\mu}\ll 1$, we work with the approximate solution
\bea
\label{x-thr}
\varepsilon \approx  \frac{3(x_{\rm thr,\mu})^{2/3}}{16 + 40(x_{\rm thr,\mu})^{2/3}}~, {\rm or}
~~ (x_{\rm thr,\mu})^{2/3} \approx \frac{16 \varepsilon}{3-40\varepsilon}~.
\eea 
Inserting this and $x=1/8-\varepsilon$ into  (\ref{S-n}) we arrive at 
\bea
\label{SU-n}
S^U_\infty(n) &=& \frac{3}{32}\left(\frac{3\pi^2 n}{2} \right)^{1/3} 
\left(1+\frac{8}{3}\varepsilon\right)\left(1-8\varepsilon\right)^{1/3}\nonumber\\
&&\times \left[1+\frac{48\varepsilon}{3-40\varepsilon}(1-8\varepsilon)^{-2/3}\right]^{1/3} 
\nonumber\\
&\simeq& \frac{3}{32}\left(\frac{3\pi^2 n}{2} \right)^{1/3} 
\left(1-\frac{64}{9}\varepsilon^2\right)
\nonumber\\
&&\times \left[1+\frac{48}{3}\varepsilon(1+\frac{40}{3}\varepsilon)(1+\frac{16}{3}\varepsilon)\right] ~,
\eea
where in the last step Taylor expansions have been used. Since $\varepsilon$ according  to (\ref{x-thr}) 
depends on the density via the muon threshold  $x_{\rm thr,\mu}$ as defined in Eq.~(\ref{mu-thr}), the 
result (\ref{SU-n}) represents a density dependent correction of the behaviour (\ref{S1}) which is obtained in
the limit $\varepsilon\to 0$.
The result (\ref{SU-n}) is shown as the dotted line in Fig.~\ref{E-DD2_large} which excellently describes 
the universal behaviour of the symmetry energy contribution to the neutron star EoS for densities above
$\sim 2~n_0$.

In a last step we recall that the symmetry energy is known exactly at the Danielewicz-Lee reference 
density $n^*$ and by construction all symmetry energies go through this point. 
Therefore, the exact value (\ref{S*-value}) of $S^*=22.04$ MeV shall be part of the USEC that we construct now by an extrapolation from the asymptotic behaviour  $S^U_\infty(n)$ through $S^*$ with an exponential ansatz
\bea
\label{SU}
S^U(n)=S^U_\infty(n) - {\rm e}^{-\alpha(n-n^*)/n^*}\left[S^U_\infty(n^*) - S^* \right],
\eea
where the slope parameter $\alpha$ regulates how fast the asymptotic solution is reached.
This analytic formula (\ref{SU}) for the USEC to the neutron star EoS is the main result of this work.
Its behaviour is illustrated in Fig.~\ref{E-DD2_large} in comparison to the contributions based on the 
DD2-type  symmetry energies.
In Fig.~\ref{E-DD2_large} we show the USEC (\ref{SU}) for the two choices of $\alpha=1.0$ and 
$\alpha=1.7$ which span the variation of the USEC in the vicinity of $n^*$ due to variations of the symmetry energy functional. 
At larger densities $n\gsim 3~n_0$, these variations become unimportant and the asymptotic behaviour of is reached.

\section{Conclusions}

We have developed general relationships for the nuclear symmetry energy and their characterizing parameters $J$, $L$ and $K_{\rm sym}$ in the vicinity of the saturation density that are based on the knowledge of the value $E_s^*$ for the symmetry energy from IAS at the density $n^*\simeq 7/10~n_0$.

We have investigated two examples for parametrizations of the high-density behaviour of the symmmetry energy: a MDI-type form and a DD2-type form. They are equivalent in the region of the saturation density where the parameters are defined and both are fixed to the Danielewicz-Lee point at $n=n^*$.

The proximity of $n^*$ and $n_0$ allows to define the symmetry energy parameters in terms of the results of the IAS analysis by Danielewicz and Lee. Both types of parametrizations for the generic high-density behaviour of the symmetry energy show a universal response to parameter changes in the region of the saturation density and result in the same $L-J$ correlation.  

The resulting consequences for neutron star matter are twofold.
Bulk properties of neutron stars which very much depend on integral quantities can be given as functions 
of quantities at the representative mean baryon density which for stars of the quite typical mass 
$1.3~M_\odot$ turns out to be equal to the saturation density for the DD2 EoS.   
These properties are insensitive to the specific form of the high-density behaviour of the symmetry energy.

Quantities which depend on the fact whether a threshold density is passed or not are very sensitive to the high-density behaviour of $E_{s}(n)$ and as an example we consider the direct Urca process.
It is remarkable that the critical proton fraction for the onset of the direct Urca process turns out to be a good
measure for the discrimination between two types of high-density behaviour: as long as the DU process is not switched on the symmetry energy contribution to the neutron star EoS behaves in a universal fashion!

We have derived a general formula for the universal high-density behaviour of this contribution which applies for a rich class of symmetry energy functions including the DD2-type ones and the APR EoS.
We present an extrapolation of the analytic formula from high to low densities, joining the asymptotic solution with the exact value at the IAS reference density.

The result of our study allows to extract the symmetric part of the nuclear EoS from NS phenomenology, e.g., from a measurement of the M-R relationship.
Vice-versa, the measurement of the high-density behaviour of the symmetric EoS, e.g., in heavy-ion collision experiments would allow to predict the EoS and thus the M-R relation for neutron  stars.

\subsection*{Acknowledgements}
We acknowledge discussions with Jim Lattimer in an early stage of this work and with Pawel Danielewicz during his recent visit at University of Wroclaw. 
We are grateful to Gerd R\"opke, Stefan Typel and Hermann Wolter for careful reading and detailed comments on the manuscript. 
We thank Stefan Typel for providing the EoS data for the DD2 models. 
This work was supported by the Polish National Science Center (NCN) under grant No. 
UMO-2014/13/B/ST9/02621 (D.E.A-C. and D.B.) and  
UMO-2013/09/B/ST2/01560 (T.K.).


\begin{thebibliography}{99}
\bibitem{footnote}
  In order to avoid complications with the muon threshold
that actually is very close to the reference density $n^*$ anyway,
we may take the freedom to choose a muon mass that would lead to a coincidence of muon threshold and reference density ($n_{\rm thr,\mu}=n^*$) entailing $m_\mu^*=(3\pi^2 n^* x^*)^{1/3}=96~$MeV. 
Compared to the physical muon mass $m_\mu=105.6~$MeV, this 10\% error is acceptable for the sake 
of clarity of our arguments.
\end{thebibliography}
\end{document}